\documentclass[aps,preprint,a4paper,showpacs,showkeys,superscriptaddress]{revtex4-1}
\usepackage{latexsym}
\usepackage{amsmath,amssymb}
\usepackage{graphicx}
\usepackage{subfigure}
\usepackage{xcolor}
\usepackage{cancel}

\usepackage{hyperref}     % color hypertext for pdflatex
\hypersetup{colorlinks,%
  citecolor=blue,%
  linkcolor=cyan,%
  pdftex}

\begin{document}

%\preprint{arXiv:yymm.nnnn [gr-qc]}

\title{Validity of black hole complementarity in the BTZ black hole}

\author{Yongwan Gim}%
\email[]{yongwan89@sogang.ac.kr}%
\affiliation{Department of Physics, Sogang University, Seoul, 04107,
  Republic of Korea}%
\affiliation{Research Institute for Basic Science, Sogang University,
  Seoul, 04107, Republic of Korea} %

\author{Wontae Kim}%
\email[]{wtkim@sogang.ac.kr}%
\affiliation{Department of Physics, Sogang University, Seoul, 04107,
  Republic of Korea}%

\date{\today}

\begin{abstract}
Based on the gedanken experiment for black hole complementarity
in the Schwarzschild black hole, we calculate the energy
required to duplicate information in the BTZ black hole
under the assumption of absorbing boundary condition
and its dual solution of the black string, respectively,
in order to justify the validity of the no-cloning theorem in quantum mechanics.
For the BTZ black hole,
the required energy for the duplication of information can be made fairly small,
whereas for the black string it exceeds the total mass of the black string,
although they are related to each other under the dual transformation.
So, the duplication of information might be possible in the BTZ black hole
in contrast to the case of the black string, so that the no-cloning theorem could be violated for the former case.
To save the duplication of information for the BTZ black hole,
we perform an improved gedanken experiment by using the {\it local} thermodynamic quantities near the
horizon rather than those defined at infinity, and show that the no-cloning theorem
could be made valid even in the BTZ black hole.
We also discuss how this local treatment for the no-cloning theorem can be applied to the black string as well as the Schwarzschild black hole innocuously.
\end{abstract}

% \pacs{04.70.Dy, 04.62.+v, 04.60.Kz }

%\keywords{Hawking Radiation, Trace anomaly, Tolman temperature}

\maketitle
%%%%%%%%%%%%%%
%% RevTeX Style End %%
%%%%%%%%%%%%%%

%\newcommand{\lp}{\ell_P}

\section{Introduction}
\label{sec:intro}

%Information loss problem and Hawking radiation and Page time
The evaporation of black holes \cite{Hawking:1974rv, Hawking:1974sw}
would lead to the information loss paradox \cite{Hawking:1976ra},
where the black holes formed by the collapse of quantum state would eventually disappear completely
 and the black hole information would be lost.
However, this problem could be solved for the distant observer outside the horizon
by assuming that the Hawking radiation would carry the black hole information.
In this assumption, the fixed observer (Bob)
could gather the information of the infalling matter state
through the Hawking radiation
after the Page time when the black hole has emitted half of its initial Bekenstein-Hawking entropy
and
the information of the black hole starts to be emitted by the Hawking radiation \cite{Page:1993wv}.
%Duplication of information
If the infalling observer (Alice) could send the message to Bob jumped into the black hole,
then Bob would have the duplicated information,
which is the violation of the unitarity
in quantum theory.

%Complementarity
The above information cloning problem arises obviously
when one attempts to correlate the experimental results performed on both sides of the horizon.
So, it has been proposed that black hole complementarity (BHC) as a solution to this paradox
should reconcile general relativity and quantum mechanics \cite{Susskind:1993if, Susskind:1993mu, Stephens:1993an}.
It means that such a paradox never occurs
since the observer inside the horizon is not in the causal past of
any observer who measures the information through the Hawking radiation
outside the horizon  \cite{Susskind:1993if, Susskind:1993mu}.
%Gedanken experiment
A specific gedanken experiment on the Schwarzschild black hole
indicates that the required energy to correlate the observations of both sides of the horizon
exceeds the mass of the black hole \cite{Susskind:1993mu}.
In other words,
the information must be encoded into the message with super-Planckian frequency.
Thus it turns out that the unitarity on the Schwarzschild black hole
%possessing the curvature singularity which plays a crucial role in determining the required energy
could survive nicely in virtue of BHC.

%%%%%%%%%%%%%%%%%%%%%%%%%

On the other hand, the AdS/CFT correspondence
allows us to study gravitational systems in terms of a dual gauge theory \cite{Maldacena:1997re, Gubser:1998bc, Witten:1998qj}.
The unitarity of the CFT strongly suggests that
no violations of causality should be apparent to local observers
and BHC should be valid
 during the process of black hole evaporation
from the AdS/CFT perspective \cite{Lowe:1999pk, Lowe:2009mq}.
%And one formulated such a proposal for how this may come about in Schwarzschild-AdS black hole and also discuss a version of the information loss paradox and its resolution \cite{Maldacena:2001kr}.
Now,
one might wonder how BHC works on the gravity side in asymptotically AdS black holes
such as the three-dimensional Ba${\rm \tilde{n}}$ados-Teitelboim-Zanelli (BTZ) black hole \cite{Banados:1992wn}.
At first sight, one might be tempted to expect that
BHC for the BTZ black hole is simply valid since the BTZ black hole under the dual transformation \cite{Horowitz:1993jc}
is related to the asymptotically flat black string solution
where the geometric structure is similar to that of the Schwarzschild black hole.

If the usual reflective boundary condition is chosen for asymptotically AdS black holes,
 the Hawking radiation from the black hole in the bulk can be reflected back into it,
 so that
the black hole does not evaporate and
remains in  thermal equilibrium with its Hawking radiation \cite{Hawking:1982dh}.
% that gets reflected back from infinity.
 However,
one could choose an absorbing boundary condition by coupling a bulk scalar field representing the Hawking radiation
to an auxiliary field at the boundary of AdS  \cite{Rocha:2008fe}.
Since such a coupling permits energy to be transferred from the bulk field to the auxiliary field,
 asymptotically AdS black holes take only a finite amount of time to evaporate away
even though its initial mass is arbitrarily large
 \cite{Rocha:2009xy, Rocha:2010zz, Ong:2015fha, Engelsoy:2016xyb,Page:2015rxa}.
It is worth noting that in the semiclassical approximations
 the coupling of the fields at the boundary does not affect the geometry of AdS  \cite{Almheiri:2013hfa}.
By assuming the absorbing boundary condition at infinity,
 %corresponding to zero incoming flux
the information loss paradox \cite{Lowe:1999pk},
BHC \cite{Lowe:2009mq}, and firewalls \cite{Almheiri:2013hfa}
have been studied  in asymptotically AdS black holes.
According to these arguments,
we will assume the absorbing boundary condition at infinity
in order to discuss BHC in the BTZ black hole.

In this paper, we shall calculate the energy
required to duplicate information on the static BTZ black hole with the absorbing boundary condition
and its dual solution of the neutral black string, respectively,
in order to justify whether the no-cloning theorem in quantum mechanics could be valid
 or not at both sides.
We shall find that
the required energy for the duplication of information in the BTZ black hole can be made very small,
whereas the required energy for the black string exceeds the super-Planckian scale.
Consequently, it means that
the duplication of information might be possible for the BTZ black hole
in contrast to the case of the black string so that the no-cloning theorem could be violated for the former case.
To evade the duplication of information for the BTZ black hole,
we perform the gedanken experiment by using the {\it local} thermodynamic quantities
defined near the horizon rather than those defined at infinity, and
show that the no-cloning theorem could be valid even in the
BTZ black hole.
Finally, we discuss this local treatment
working well for the black string as well as the Schwarzschild black hole
without any modification of standard conclusion.
In Sec.~\ref{sec:Sch}, we recapitulate the well-established gedanken experiment
to determine the required energy for cloning the infalling information on the Schwarzschild black hole along the line of Ref.~\cite{Susskind:1993mu}.
In Sec.~\ref{sec:three},
we calculate the required energy for cloning and then point out the violation of the no-cloning theorem in the BTZ black hole.
In Sec.~\ref{sec:boundary}, we revisit the above gedanken experiment
by employing the local thermodynamic quantities for the BTZ black hole, and then resolve the violation of the no-cloning theorem.
Finally, conclusion and discussion will be given in Sec.~\ref{sec:con}.

\section{BHC in the Schwarzschild black hole}
\label{sec:Sch}

Let us encapsulate the gedanken experiment presented in Ref.~\cite{Susskind:1993mu}.
The two observers, Alice and Bob, are hovering outside the Schwarzschild black hole
 described by
the length element as
%\begin{equation}\label{eq:}
%ds^2 = -\left(1-\frac{2M}{r}\right) dt^2 + \frac{1}{1-\frac{2M}{r}} dr^2 + r^2 d\Omega^2
%\end{equation}
\begin{equation}\label{eq:Kruskal1}
ds^2=-\frac{32 M^3}{r} e^{-\frac{r}{2M}}dUdV + r^2 d\Omega^2,
\end{equation}
where $U=\pm e^{-\frac{(t-r^*)}{4M}}$,
$V=e^{\frac{(t+r^*)}{4M}}$, and $r^*=r+2M \ln \left(|r-2M|/2M\right)$,
and the constants are $G=\hbar =c=k_{\rm B}=1$.
The plus and minus signs in $U$ coordinate are for the inside and outside the horizon,
respectively.
To get the Page time for the old black hole  \cite{Page:1993wv},
we consider the four-dimensional Stefan-Boltzmann law in the asymptotically flat spacetime as
\begin{equation}\label{eq:SB1}
\frac{d {\cal E}}{dt}=-A \sigma T^4,
\end{equation}
where ${\cal E}, A, T$ are the internal energy, area, temperature of a black body, and $\sigma$ denotes the Stefan-Boltzmann constant.
Now, the internal energy $E$, the area $A$, and the temperature $T$ are
identified with ${\cal E}=M,~A=16 \pi M^2,$ and $T=1/(8\pi M)$ for the Schwarzschild black hole.
Then, the Page time $t_{\rm P}$ when the initial Bekenstein-Hawking entropy shrinks in half
can be easily calculated from the Stefan-Boltzmann law \eqref{eq:SB1} as
\begin{equation}\label{eq:pageT}
t_{\rm P} \sim M^3.
\end{equation}

%%%%%%%%%%%%%%%%%%%%%%%%%%%%%
% Fig: %%
%%%%%%%%%%%%%%%%%%%%%%%%%%%%%
\begin{figure}[hpt]
  \begin{center}
  \includegraphics[width=0.5\linewidth]{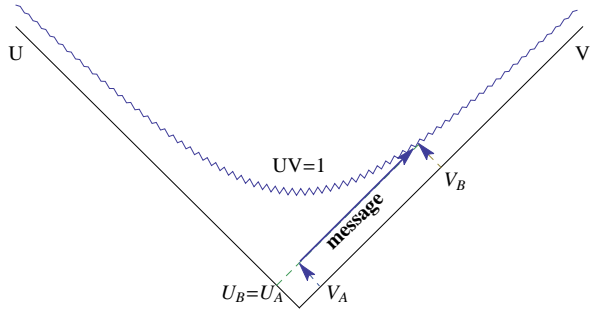}
  \end{center}
  \caption{For the Schwarzschild black hole (and the black string, BTZ black hole),
  the origin is denoted by the wiggly curve where $UV=1$ in terms of the Kruskal-Szekeres coordinates.
  Alice is passing through the horizon at $V_{\rm A}$, and Bob will jump into the horizon at $V_{\rm B}$ after the Page time.
 Alice should send the message to Bob at least at $U_{\rm A}$ before Bob hits the singularity.}
  \label{fig:penrose1}
\end{figure}
%%%%%%%%%%%%%%%%%%%%%%%%%%%%%

Next, let us suppose that Alice passes through the horizon at $V_{\rm A}$ in Fig. \ref{fig:penrose1},
and after the Page time $t_{\rm P}$ Bob falls through the horizon with carrying a record of his measurements at $V_{\rm B}$.
To receive the message from Alice before Bob hits the curvature singularity at $UV=1$,
Alice should send the message to Bob before she reaches
$U_{\rm A}=U_{\rm B}=V_{\rm B}^{-1}= e^{-t_{\rm P}/(4M)} \sim e^{-M^2}$, where $r=0$.
So, the proper time $\Delta \tau$ for Alice to send the message before passing through $U_{\rm A}$
can be calculated from Eq. \eqref{eq:Kruskal1} near the horizon $r\approx r_{\rm H}$ as
\begin{align}
\Delta \tau^2 \approx \frac{32M^3}{r_{\rm H}}e^{-\frac{r_{\rm H}}{2M}} (U_{\rm A}-0) \Delta V_{\rm A} \sim M^2 e^{-M^2}, \label{eq:tau}
\end{align}
where $\Delta V_{\rm A}$ is a nonvanishing finite value near $V_{\rm A}$ for the free-fall \cite{Susskind:1993mu}.
Then, the required energy $\Delta E$
is calculated as
\begin{equation}\label{eq:DE}
\Delta E \sim \frac{1}{M} e^{M^2},
\end{equation}
which definitely exceeds the black hole mass, {\it i.e.}, $\Delta E \gg M$.
It dictates that the message must be encoded into radiation with super-Planckian frequency.
Therefore, the no-cloning theorem is valid in such a way that BHC is complete.

Note that the Page time was calculated at asymptotic infinity,
even though Bob should be staying near the horizon in the gedanken experiment \cite{Susskind:1993mu}.
Since the time interval $\Delta t_{\rm B}$ of Bob is always longer than the interval $\Delta t_{\infty}$ of the asymptotic observer thanks to time dilation
% $\Delta t_{\rm B} > \Delta t_{\infty}$,
for the asymptotic flat black holes,
the required energy calculated from $\Delta t_{\infty}$  is also larger than that from  $\Delta t_{\rm B}$. However, for asymptotically non-flat black holes, the above gedanken experiment would be
non-trivial, which is the main theme of the present work.

\section{No-cloning theorem in the black string and the BTZ black hole}
\label{sec:three}

Let us first mention the dual transformation between the BTZ black hole and
the black string \cite{Horowitz:1993jc, Horowitz:1993wt}
by using the three-dimensional low energy string action defined by
\begin{equation}\label{eq:action}
S=\frac{1}{2\pi}\int d^3x\sqrt{-g}e^{-2\phi}\left[ \frac{4}{\ell^2} +R+4(\nabla \phi)^2-\frac{1}{12}H_{\mu\nu\rho}H^{\mu\nu\rho}\right],
\end{equation}
where $\phi$ is the dilaton field, $H_{\mu\nu\rho}$ satisfying $H=dB$ is the three-form field strength, and
the cosmological constant is $\Lambda =-{\ell^{-2}}$.
The equations of motion for the action \eqref{eq:action} are given by
\begin{align}
&R_{\mu\nu}+2\nabla_\mu \nabla_\nu \phi -\frac{1}{4}H_{\mu\lambda\rho}H_{\nu}^{\lambda\rho} =0, \notag \\
&\nabla_\mu(e^{-2\phi} H_{\mu\lambda\rho} )=0, \label{eq:EOM} \\
&R-\frac{1}{12}H^2+4\left(\nabla^2 \phi - (\nabla \phi)^2 +\frac{1}{\ell^2}\right)=0 \notag
\end{align}
with respect to each field.
One of the solutions to Eq.~\eqref{eq:EOM} is the static BTZ black hole described by the
metric,
\begin{equation}\label{eq:BTZ}
ds^2=-\left(-M+\frac{r^2}{\ell^2}\right)dt^2 + \frac{1}{-M+\frac{r^2}{\ell^2}}dr^2+r^2d\varphi^2
\end{equation}
with $B_{\varphi t}=r^2 /\ell^2$, $\phi=0$, where $\ell$ is the radius of AdS space.
The metric solution \eqref{eq:BTZ} with  $\{B, \phi\}$
is independent of the coordinate $\varphi$, which gives
the dual solution through the dual transformation  as \cite{Buscher:1987sk, Buscher:1987qj}
\begin{align}
&\tilde{g}_{\varphi\varphi} =\frac{1}{g_{\varphi \varphi}},\qquad
\tilde{g}_{\varphi \mu}=\frac{B_{\varphi \mu}}{g_{\varphi\varphi}}, \label{eq:}\\
&\tilde{g}_{\mu\nu}=g_{\mu \nu} - (g_{\varphi \mu} g_{\varphi \nu}-B_{\varphi \mu}B_{\varphi \nu})/g_{\varphi\varphi},\\
&\tilde{B}_{\varphi \mu} =\frac{g_{\varphi \mu}}{g_{\varphi \varphi}},\qquad
\tilde{B}_{\mu\nu}=B_{\mu\nu}-\frac{2}{g_{\varphi\varphi}}g_{\varphi [\mu}g_{\nu]\varphi}, \\
&\tilde{\phi}=\phi-\frac{1}{2}\ln g_{\varphi\varphi},
 \end{align}
where $\mu,~\nu$ run over all coordinates except $\varphi$.
The dual solution is obtained by applying the dual transformation to a translational symmetry in the direction $\varphi$,
\begin{equation}\label{eq:}
d\tilde{s}^2=Mdt^2+\frac{2}{\ell}dtd\varphi + \frac{1}{r^2}d\varphi^2 + \frac{1}{-M+\frac{r^2}{\ell^2}}dr^2,
\end{equation}
and $\tilde{B}_{\varphi t}=0,~\tilde{\phi}=-\ln r$.
After a diagonalization
through $t=(\hat{x}-\hat{t})/\sqrt{M},~ \varphi=\ell\sqrt{M}\hat{t}$, and $ r^2=\ell \hat{r},$
the neutral black string solution is finally obtained as
\begin{align}\label{eq:BS}
d\tilde{s}^2 =-\left(1-\frac{\cal M}{\hat{r}}\right)d\hat{t}^2+ \frac{\ell^2}{4\hat{r}^2} \frac{1}{1-\frac{\cal M}{\hat{r}}}d\hat{r}^2+d\hat{x}^2
\end{align}
with $\tilde{B}_{\hat{x}\hat{t}}=0,~\tilde{\phi}=-\frac{1}{2}\ln(\hat{r}\ell)$,
where the ADM mass per unit length of the black string is ${\cal M} =\ell M$.
The BTZ black hole \eqref{eq:BTZ} and the black string \eqref{eq:BS}
are the solutions to the equations of motion \eqref{eq:EOM} and
they have the same entropy and mass \cite{Horowitz:1993wt, Ho:1997uk}; however,
the geometric properties are very different.

Let us now investigate the validity of BHC
in the black string whose metric \eqref{eq:BS} is rewritten as
\begin{align}\label{eq:BS2}
d\tilde{s}^2 =- \frac{\ell^2 \cal{M}}{\hat{r}}d\hat{U}d\hat{V}+d\hat{x}^2
\end{align}
 in terms of the Kruskal-Szekeres coordinates defined by
$\hat{U}= \pm e^{-(\hat{t}-\hat{r}^*)/\ell},~\hat{V}=e^{(\hat{t}+\hat{r}^*)/\ell}$, and $\hat{r}^*=\frac{\ell}{2} \ln\left(|\hat{r}-{\cal M}|/{\cal M}\right)$.
To calculate the Page time in three dimensions,
we use the three-dimensional Stefan-Boltzmann law
written as \cite{Landsberg:1989}
\begin{equation}\label{eq:SB2}
\frac{d{\cal E}}{dt} = -A \sigma T^3,
\end{equation}
where ${\cal E}, A, T$ are the mass,
area, and Hawking temperature of the black string, respectively, and
$\sigma$ is the three-dimensional Stefan-Boltzmann constant.
The conserved charge $\tilde{{\cal E}}$ for a time-like killing vector,
the area $\tilde{A}_{\rm H}$, and the Hawking temperature $\tilde{T}_{\rm H}$  of the black string
are calculated as $\tilde{{\cal E}}=2\sqrt{{\cal M}/\ell}$,
$\tilde{A}_{\rm H}=\int d\hat{x}=2\pi/\sqrt{\ell {\cal M}},$ and $\tilde{T}_{\rm H}=1/(2\pi \ell)$ \cite{Horowitz:1993wt},
and the Wald entropy is given as $\tilde{S}_{\rm W}=4\pi \sqrt{\ell {\cal M}}$ \cite{Wald:1993nt}.
By plugging them into Eq. \eqref{eq:SB2}, the Page time $\hat{t}_{\rm P}$
is calculated as
\begin{equation}\label{eq:tp}
\hat{t}_{\rm P}\sim \ell^3 {\cal M}.
\end{equation}

Let us suppose that Alice sends the message to Bob before Bob hits the singularity
as shown in Fig. \ref{fig:penrose1}.
Since the black string has the curvature singularity at $\hat{r}=0$ or
$\hat{U}\hat{V}=1$,
$\hat{U}_{\rm A}$ is written as
$\hat{U}_{\rm A}=\hat{U}_{\rm B}=\hat{V}_{\rm B}^{-1}= e^{-\hat{t}_{\rm P}/\ell} \sim e^{-\ell^2 {\cal M}^2}$
for the Page time \eqref{eq:tp}.
From the metric \eqref{eq:BS2}, the proper time is calculated as
\begin{align}
\Delta \tilde{\tau}^2 &= \frac{\ell^2 {\cal M}}{\hat{r}_{\rm H}}\hat{U}_{\rm A} \Delta \hat{V}_{\rm A}  \notag \\
&\sim \ell^2 e^{-\ell^2 {\cal M}}
\end{align}
near the horizon.
So the required energy is immediately read off from the Heisenberg uncertainty principle as
\begin{equation}\label{eq:DeltaE}
\Delta \tilde{E} \sim \frac{1}{\ell} e^{\ell^2 {\cal M}},
\end{equation}
which dictates that the message must be encoded into radiation with super-Planckian frequency.
In other words, for the large black string, the required energy for cloning exceeds the total mass of the black string, {\it i.e.},
$\Delta \tilde{E} \gg \tilde{{\cal E}}=2\sqrt{{\cal M}/\ell}$.
The analysis of the gedanken experiment on the black string
respects the no-cloning theorem in the principle of quantum mechanics,
 so that BHC is still valid.

%We proceed with the similar gedanken experiment on the BTZ black hole.
Before proceeding with a similar gedanken experiment for the BTZ black hole,
let us comment on evaporation of the BTZ black hole.
Actually, one could easily discuss the information loss paradox
if the black hole evaporated.
%In many aspects, AdS behaves just like a box.
If we place a small enough black hole in AdS,
it will evaporate before filling the surrounding space.
However, a large black hole does not evaporate completely and instead reaches a configuration of thermal equilibrium
in contrast with the small black hole,
since the reflecting boundary condition of asymptotically AdS black holes prevents such black holes from evaporating completely
\cite{Hawking:1982dh, Cai:2007vv}, and
this kind of feature can also be found in BTZ black hole \cite{Eune:2013qs}.
In this respect, the information loss paradox should be studied
by assuming the evolving geometry of the asymptotically AdS black holes \cite{Lowe:1999pk}.
%in order to address the information loss paradox,
To support this argument,
the  boundary was required to become partially absorptive
by allowing energy to be transferred between bulk fields and external fields \cite{Rocha:2008fe}.
It turns out that asymptotically AdS black holes take only a finite amount of time to evaporate away
even though its initial mass is arbitrarily large.
The relevant discussions in the bulk and the AdS boundary in connection with the evaporating AdS black holes with the absorptive boundary condition could be found in a few literatures
 \cite{Rocha:2009xy, Rocha:2010zz, Page:2015rxa, Ong:2015fha, Engelsoy:2016xyb}.
%And recently,  this argument is extended more generic black holes in AdS \cite{Ong:2015fha}.
It is worth mentioning that
 the coupling of the fields at the boundary does not affect the geometry of AdS under the semiclassical approximations \cite{Almheiri:2013hfa}.
From now on, we will assume the absorbing boundary condition at infinity
in order to discuss the no-cloning theorem for BTZ black hole.

Let us start with the geometry of the BTZ black hole described by the metric \eqref{eq:BTZ}
written as
\begin{equation}\label{eq:BTZ2}
ds^2=-\frac{\ell^2}{r_{\rm H}^2}(r+r_{\rm H})^2 dU dV+r^2d\varphi^2
\end{equation}
in the Kruskal-Szekeres coordinates, where
$U= \pm e^{-\frac{r_{\rm H}}{\ell^2}(t-r^*)},~V=e^{\frac{r_{\rm H}}{\ell^2}(t+r^*)}$,
and $r^*=\ell^2/(2r_{\rm H})\ln(|r-r_{\rm H}|/(r+r_{\rm H}))$,
and $r_{\rm H}=\ell \sqrt{M}$.
Plugging the ADM mass ${\cal E}=M$, the area $A_{\rm H}=2\pi \ell \sqrt{M}$,
and the Hawking temperature
\begin{equation}\label{eq:THBTZ}
T_{\rm H}=\sqrt{M}/(2\pi \ell)
\end{equation}
into the three-dimensional Stefan-Boltzmann law \eqref{eq:SB2},
we obtain the Page time $t_{\rm P}$
as
%with Wald entropy $S_{\rm W}=4\pi\ell\sqrt{M}$.
\begin{equation}\label{eq:tp2}
t_{\rm P} \sim \frac{\ell^2}{M}.
\end{equation}

Note that
the static BTZ black hole is geodesically incomplete \cite{Cruz:1994ir}.
If the geometry of the static BTZ black hole is extended to
$r=0$, the distributional curvature scalar $R$  with a possible distributional source is found
to be $R=\pi (1+M)\delta^{(2)}_{(0)}(r^2)-6/\ell^2$ \cite{Pantoja:2002nw},
where it reduces the curvature of the pure AdS spacetime of $R=-6/\ell^2$ for $M=-1$.
For the static BTZ black hole, the curvature singularity and the point source as the mass of the BTZ black hole are placed at $r=0$, {\it i.e.}, $UV=1$.
From the Page time \eqref{eq:tp2}, the critical point $U_{\rm A}$ for Alice can be obtained
as $U_{\rm A}=U_{\rm B}=V_{\rm B}^{-1}= e^{-(r_{\rm H}/\ell^2)t_{\rm P}} \sim e^{-\ell/\sqrt{M}}$.
Then, from the metric \eqref{eq:BTZ2},
the proper time $\Delta \tau$ for Alice is obtained as
\begin{align}\label{eq:}
\Delta \tau \sim \ell e^{-\frac{\ell}{\sqrt{M}}}
\end{align}
near the horizon.
Eventually, the required energy is given as
\begin{equation}\label{eq:BTZE}
\Delta E \sim \frac{1}{\ell} e^{\frac{\ell}{\sqrt{M}}}.
\end{equation}
Unlike the Schwarzschild black hole or the black string,
the required energy \eqref{eq:BTZE} for duplication of information can be made small compared to the ADM mass of the black hole,
 $\Delta E \ll M$, for the large BTZ black hole.
It means that the no-cloning theorem of quantum information can be violated,
so that BHC appears to be invalid. We will show that this is not the case in the next section.

\section{One resolution for the BTZ black hole}
\label{sec:boundary}
The violation of the no-cloning theorem for the BTZ black hole is problematic, since
there is no reason for the unitary quantum theory to be ill-defined
when gravitational effects are considered in the semi-classical regime.
To evade the duplication of information in the BTZ black hole,
we will try to perform the gedanken experiment by employing the local thermodynamic quantities
defined in the  near horizon rather than those defined at infinity,
since at a finite distance from the horizon
Bob is gathering information from the Hawking radiation.
Measurement of the Hawing radiation carrying information
will be done in the local inertial frame first and then it will be transformed to Bob in
the accelerating frame, because there is no asymptotically flat region for the BTZ black hole.

Let us obtain the local thermodynamic quantities
for a local observer at a radius of $r_{\rm B}$ of the black hole.
The local temperature $T_{\rm loc}$ can be
obtained from the Tolman's law \cite{Tolman:1930zza, Tolman:1930ona},
\begin{equation}\label{eq:Tloc}
T_{\rm loc}=\frac{T_{\rm H}}{\sqrt{-g_{tt}(r_{\rm B})}}.
\end{equation}
And the thermodynamic internal
energy ${\cal E}_{\rm loc}$ can be computed from the first law of black hole thermodynamics as
 $d{\cal E}_{\rm loc}=T_{\rm loc}dS$ where
$B_{\varphi t}=r^2 /\ell^2$, $\phi=0$, which yields
${\cal E}_{\rm loc} = {\cal E}_0 -2 \sqrt{-M+r_{\rm B}^2/\ell^2}$,
where ${\cal E}_{\rm loc}$ would vanish for the zero mass for ${\cal E}_0 = 2 r_{\rm B}/\ell$.

Now, let us assume that
Bob is gathering the information encoded in the Hawking radiation
near the horizon, $r_{\rm B} = r_{\rm H} +\epsilon$ where
$\epsilon$ is a finite value.
By the way,
in the local inertial frame at the position of Bob, i.e., $r=r_{\rm B}$, we can
consider the Stefan-Boltzmann law described by the local thermodynamic quantities as
\begin{equation}\label{eq:SBloc}
\frac{d{\cal E}_{\rm loc}}{d\tau_{\rm B}}=-A \sigma T_{\rm loc}^3,
\end{equation}
where
$\tau_{\rm B}$ is a proper time at the position of $r_{\rm B}$.
%and ${\cal E}_{\rm loc}$ and $T_{\rm loc}$ mean the thermal energy and the temperature of the black hole defined by a local inertial observer.
Plugging the first law of thermodynamics of $d{\cal E}_{\rm loc}=T_{\rm loc}dS$ into Eq. \eqref{eq:SBloc},
one can get the Page time measured
in the local inertial frame as
\begin{align}
 \tau_{\rm B}-\tau_0 &=-\int^{\frac{S}{2}}_{S} \frac{1}{A \sigma T_{\rm loc}^2}dS \notag \\
&= \int^{\frac{S}{2}}_{S} \frac{g_{tt}(r_{\rm B})}{A \sigma T_{\rm H}^2}dS,\label{eq:tploc}
\end{align}
which results in
\begin{equation}\label{eq:tpBTZloc}
 \tau_{\rm B}-\tau_0 \sim \ell^2 + \ell \sqrt{M} \epsilon ,
\end{equation}
where $\tau_0$ denotes the proper time when the Alice jumped into the horizon.
By applying the time dilation near the horizon due to the acceleration of the Bob's frame,
\begin{equation}
\Delta \tau_{\rm B} = \sqrt{-g_{tt} (r_{\rm B})} \Delta t_{\rm P},
\end{equation}
the Page time $t_{\rm P}$ measured by Bob who is the fixed observer outside the horizon is obtained as
\begin{equation}\label{eq:}
 t_{\rm P}  \sim \sqrt{\frac{\ell^{3} \epsilon}{2}}M^{\frac{1}{4}}+\sqrt{\frac{\ell}{2\epsilon}}\left( \ell^2 +\frac{3}{4}\epsilon^2\right) M^{-\frac{1}{4}}+\mathcal{O}\left(M^{-\frac{3}{4}}\right).
\end{equation}

Next, the critical point $U_{\rm A}$ for Alice can be given
by $U_{\rm A}=U_{\rm B}=V_{\rm B}^{-1}= e^{-(r_{\rm H}/\ell^2)t_{\rm P}} $,
so that the proper time for Alice is obtained as
\begin{equation}\label{eq:}
\Delta \tau \sim \ell e^{-\left(\sqrt{\frac{\ell \epsilon}{2}}M^{\frac{3}{4}}+\frac{1}{\sqrt{2\epsilon \ell}}\left( \ell^2 +\frac{3}{4}\epsilon^2\right) M^{\frac{1}{4}}+\mathcal{O}\left(M^{-\frac{1}{4}}\right)\right)},
\end{equation}
and the required energy from the uncertainty principle is eventually written as
\begin{equation}\label{eq:DEloc}
\Delta E \sim \frac{1}{\ell} e^{\sqrt{\frac{\ell \epsilon}{2}}M^{\frac{3}{4}}+\frac{1}{\sqrt{2\epsilon \ell}}\left( \ell^2 +\frac{3}{4}\epsilon^2  \right) M^{\frac{1}{4}}+\mathcal{O}\left(M^{-\frac{1}{4}}\right)}.
\end{equation}
For the large BTZ black hole, the required energy \eqref{eq:DEloc} definitely exceeds the mass of the black hole, ${\it i.e.},~\Delta E \gg M$,
which means that the message must be sent in quanta of energy far beyond the Planck scale.
Therefore,
in the BTZ black hole,
the no-cloning theorem of the quantum theory could be made
valid like the case of the black string as well as the Schwarzschild black hole,
so that BHC becomes complete.

%
%
%One could put the observer at asymptotic infinity.
%Then, since the local temperature \eqref{eq:Tloc} of the BTZ black hole
%vanishes,
%the Stefan-Boltzmann law  \eqref{eq:SB2} is rewritten as
%\begin{equation}\label{eq:}
%\frac{d{\cal E}}{dt} = 0.
%\end{equation}
%So, the Page time is infinity as $\Delta t_{\rm p} \rightarrow \infty$,
%which means that
%the exterior observer at asymptotic infinity never has access to enough of the quantum state to perform
%any useful measurement on the Hawking radiation.
%%Bob has to wait indefinitely outside the black hole to gather the Alice's information
%Therefore, the no-cloning theorem in the context of BHC is always valid.
%

\section{Conclusion and discussion}
\label{sec:con}

We examined the validity of BHC for the BTZ black hole
by calculating
the required energy for Alice to send the message to Bob.
From the naive gedanken experiment in section \ref{sec:three},
the energy required to duplicate Alice's information could be made small enough to duplicate it,
although the duplication of information did not occur
in the black string.
To save the violation of the no-cloning theorem for the BTZ black hole with
the absorptive boundary,
we calculated the Page time by using the local thermodynamic quantities
in the Stefan-Boltzmann law and then transformed it to the fixed frame,
because there is no asymptotic region in the BTZ black hole.
Consequently, we found that
BHC is still valid for the BTZ black hole.

Precisely, one might wonder
what happens if the local thermodynamic quantities are
employed in the calculations of the
required energy cloning the information in the Schwarzschild black hole and the black string.
For the case of the Schwarzschild black hole,
from the Stefan-Boltzmann law \eqref{eq:SB1} with the local temperature defined by $T_{\rm loc}=1/(8\pi M \sqrt{1-2M/r_{\rm B}})$,
the Page time is calculated as
$t_{\rm loc} \sim M^4/\epsilon+M^3$,
where $\epsilon=r_{\rm B}-2M$,
and for the case of the black string the Page time $\hat{t}_{\rm P}$
is also obtained as
$\hat{t}_{\rm loc} \sim \ell^3 {\cal M}^2/\epsilon + \ell^3 {\cal M} $.
Note that the local treatment for $\epsilon$ makes the Page time $t_{\rm loc}$ and $\hat{t}_{\rm loc}$ longer so that the required energy for cloning the information is enhanced.
Therefore, in the large black hole limit, the gedanken experiment with the local thermodynamic quantities is still consistent with the conventional result in Ref.~\cite{Susskind:1993mu}.

Now, we would like to discuss the relationship between information loss paradox and boundary conditions.
In the usual reflective boundary condition,
the external observer simultaneously receives incoming and outgoing fluxes of the Hawking radiation,
so that
the black hole does not evaporate and
remains in  thermal equilibrium.
%with its Hawking radiation.
%no net radiation is received by the exterior observer
Since the black hole does not accommodate that the initial entropy shrinks in half,
%and information will stay in the black hole forever
no bit of information would leak out from the black hole
so that there is no information loss paradox.
However,
%the information loss occurs when
in the absorptive boundary condition,
%is chosen in the asymptotically AdS black hole,
%so that the information loss problem and BHC should be studied in this case.
%It has been shown that
 the asymptotically AdS black hole can evaporate by means of the auxiliary field at the boundary \cite{Rocha:2008fe, Rocha:2009xy, Rocha:2010zz, Ong:2015fha, Engelsoy:2016xyb};
however, there is no detailed proof in the case of the BTZ black hole.
It deserves further attention.

Finally, if the black hole does not evaporate for the BTZ black hole with the reflective boundary condition, then the Stefan-Boltzmann law \eqref{eq:SBloc}
does not work. In fact, the Stefan-Boltzmann law  \eqref{eq:SBloc} is just the relation between the net flux and the temperature of the
radiating black hole.
Here, the flux-temperature relation is valid only for the evaporating black hole with the absorptive boundary condition.
Unfortunately, we did not derive the Stefan-Boltzmann law  \eqref{eq:SBloc} for the BTZ black hole
but just assumed the form which usually holds in local inertial frames.
The explicit derivation of Eq. \eqref{eq:SBloc}
by using the influx and out-flux from the quantum-mechanical energy-momentum tensor
by assuming the absorptive boundary
seems to be beyond the scope of the present paper.
We hope this issue will be addressed elsewhere.

\acknowledgments

This work was supported by
the National Research Foundation of Korea(NRF) grant funded by the
Korea government(MSIP) (2017R1A2B2006159).

%%%%%%%%%%%%%%%%%%%%%%%%%%%%%%%%%%%%%%%%%%%%%%%%
%%%%%%%%%%%%%%%             References         %%%%%%%%%%%%%%%%

%\bibliographystyle{mybib}
%\bibliographystyle{apsrev4-1} % PRD
%\bibliographystyle{model1-num-names}  % PLB with title + plb.bst파일 같은 디렉토리에 첨부
\bibliographystyle{JHEP}       %% JHEP.bst+ jhep.bst파일 같은 디렉토리에 첨부

\bibliography{references}

\end{document}